\documentclass[cernpreprint, atlasdraft=false, USenglish, texmf, orcidlogo]{atlasdoc}			
 
\usepackage{atlaspackage}
\usepackage{atlasbiblatex}
\usepackage{lscape} 
 
\usepackage{atlasphysics}

\graphicspath{{logos/}}
 
\usepackage{ANA-STDM-2019-17-PAPER-defs}

\AtlasTitle{Observation of $WZ\gamma$ Production in $pp$ Collisions at $\sqrt{s}=13$ TeV with the ATLAS Detector}

\AtlasAbstract{

This Letter reports the observation of $WZ\gamma$ production and a measurement of its cross-section using \mbox{$140.1 \pm 1.2~\textrm{fb}^{-1}$} of proton--proton collision data recorded at a center-of-mass energy of \mbox{13 TeV} by the ATLAS detector at the Large Hadron Collider.
The $WZ\gamma$ production cross-section, with both the $W$ and $Z$ bosons decaying leptonically, $pp \rightarrow WZ\gamma \rightarrow {\ell'}^\pm\nu\ell^{+}\ell^{-}\gamma$ ($\ell^{(')} = e, \mu$), is measured in a fiducial phase-space region defined such that the leptons and the photon have high transverse momentum and the photon is isolated.  The cross-section is found to be $2.01 \pm 0.30~(\textrm{stat.}) \pm 0.16~(\textrm{syst.})~\textrm{fb}$. The corresponding Standard Model predicted cross-section calculated at next-to-leading order in perturbative quantum chromodynamics and at leading order in the electroweak coupling constant is $1.50 \pm 0.06~\fb$. The observed significance of the \WZy\ signal is $6.3\sigma$, compared with an expected significance of $5.0\sigma$.

}
 
\author{The ATLAS Collaboration}

\usepackage{todonotes}
\usepackage{makecell}
\usepackage[utf8]{inputenc}
\usepackage[T1]{fontenc}
\usepackage{graphicx}
\AtlasRefCode{STDM-2019-17}
 
\AtlasJournalRef{\PRL 132 (2024) 021802}
\AtlasDOI{10.1103/PhysRevLett.132.021802}
 
\PreprintIdNumber{CERN-EP-2023-095}

\hypersetup{pdftitle={ATLAS document},pdfauthor={The ATLAS Collaboration}}
 
\begin{document}
 
\maketitle

Electroweak (EW) production of triboson states, $V\gamma\gamma$, $VV\gamma$, and $VVV$ ($V=\Wboson$ or \Zboson), in high-energy proton--proton ($pp$) collisions provides one of the primary means to probe the quartic interactions between EW gauge bosons and to carry out indirect searches for physics beyond the Standard Model (SM).
Although such studies are experimentally challenging because of the small cross-sections involved and the presence of significant background contributions, the ATLAS and CMS experiments at the Large Hadron Collider (LHC) have observed some of the relevant channels.
The ATLAS and CMS collaborations have observed \Zyy\ production at $pp$ center-of-mass energies $\sqrt{s}$ of $8\,\TeV$ and $13\,\TeV$~\cite{STDM-2014-01,STDM-2021-09,CMS-SMP-15-008,CMS-SMP-19-013} while \Wyy\ production has recently been observed by ATLAS at $\sqrt{s}$ = $13\,\TeV$~\cite{atlasWyyobservation}.
The combined production of three massive gauge bosons, $VVV$, has been observed at $\sqrt{s} = 13\,\TeV$ by CMS~\cite{CMS-SMP-19-014}, and the observation of \WWW\ production was reported by ATLAS~\cite{STDM-2019-09}, also at $\sqrt{s} = 13\,\TeV$. Recently, \WWy\ production has been observed by CMS at $\sqrt{s} = 13\,\TeV$~\cite{cmscollaborationWWyobservation}.
No evidence for \WZy\ or \ZZy\ production has yet been obtained. For these channels, only upper limits of approximately 2--4 times the predicted SM cross-section on the combined production of the \WWy\ and \WZy\ triboson states at $\sqrt{s} = 8$\,\TeV\ have been reported by the ATLAS~\cite{STDM-2016-05} and CMS~\cite{CMS-SMP-13-009} collaborations.
 
\begin{figure}[ht]
\begin{center}
\subfloat[]{\includegraphics[height=0.14\textwidth]{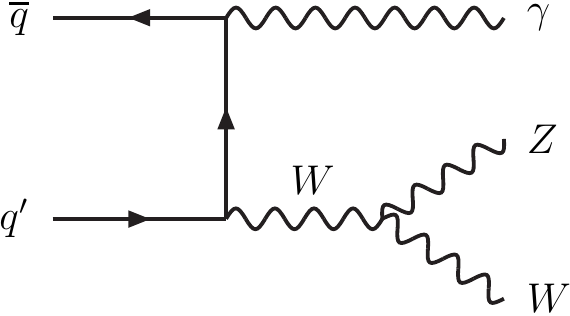}} \qquad
\subfloat[]{\includegraphics[height=0.14\textwidth]{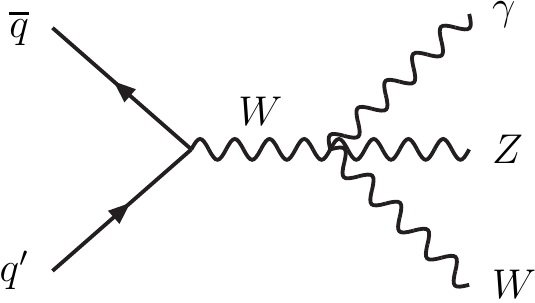}} \qquad
\subfloat[]{\includegraphics[height=0.14\textwidth]{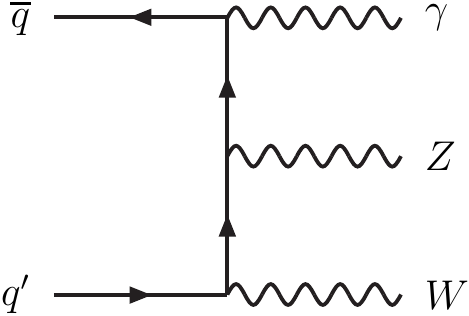}} \\  \vspace*{3mm}
\subfloat[]{\includegraphics[height=0.2\textwidth]{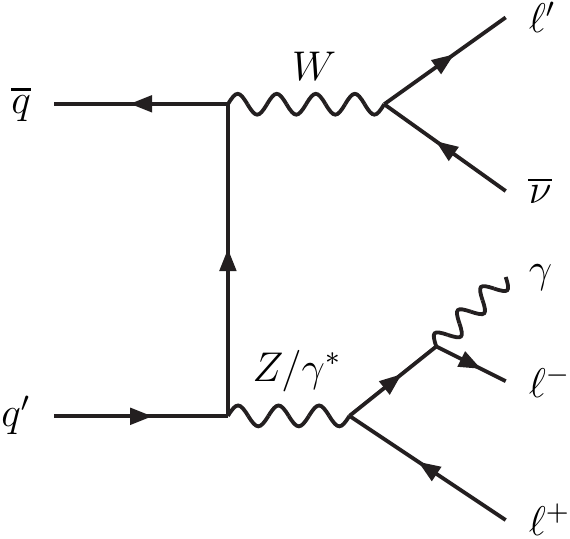}} \qquad
\subfloat[]{\includegraphics[height=0.2\textwidth]{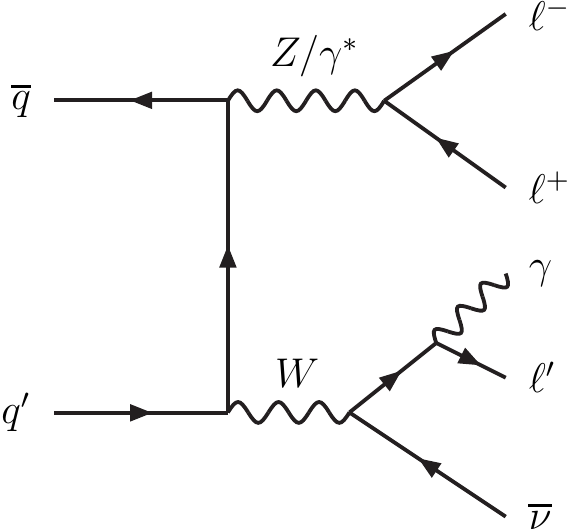}} \qquad
\caption{\small{
Representative leading-order Feynman diagrams for (a)--(c) $WZ\gamma$ production and (d)--(e) ${\ell'}^{\pm} \nu \ell^{+}\ell^{-}\gamma$ production via the FSR processes $Z\to\ell^+\ell^-\gamma$ and $W\to\ell'\nu\gamma$.
}
\label{fig:feynman}
}
\end{center}
\end{figure}

This Letter reports the observation of \WZy\ production in $pp$ collisions with both the \Wboson\ and the \Zboson\ boson decaying leptonically, $\pp \to WZ\gamma \to{\ell'}^\pm\nu\ell^{+}\ell^{-}\gamma$, where $\ell'$ and $\ell$ are an electron or a muon, using \mbox{$140.1 \pm 1.2~\textrm{fb}^{-1}$}~\cite{lumi2023, LUCID2} of data at $\sqrt{s} = 13$~\TeV\ recorded with the ATLAS detector.
The ${\ell'}^\pm\nu\ell^{+}\ell^{-}\gamma$ production cross-section is measured in a fiducial phase-space region defined such that the leptons and the photon have high transverse momentum and the photon is isolated, and including kinematic requirements which enhance the relative contribution from processes where the photon is produced directly in the initial hard-scattering interaction, as illustrated in Figures~\ref{fig:feynman}(a)--\ref{fig:feynman}(c), including the quartic interaction contribution of primary interest of Figure~\ref{fig:feynman}(b), rather than being radiated from a final-state charged lepton (final-state radiation, FSR), as illustrated in Figures~\ref{fig:feynman}(d)--\ref{fig:feynman}(e).

The ATLAS experiment~\cite{PERF-2007-01} at the LHC is a multipurpose particle detector with a forward--backward symmetric cylindrical geometry covering nearly the entire solid angle around the collision point.\footnote{ATLAS uses a right-handed coordinate system with its origin at the nominal interaction point (IP) in the center of the detector and the \(z\)-axis along the beam pipe.
The \(x\)-axis points from the IP to the center of the LHC ring, and the \(y\)-axis points upwards.
Cylindrical coordinates \((r,\phi)\) are used in the transverse plane, \(\phi\) being the azimuthal angle around the \(z\)-axis.
The pseudorapidity is defined in terms of the polar angle \(\theta\) as \(\eta = -\ln \tan(\theta/2)\).
Momentum in the transverse plane is denoted by \pT.}
Its major components are an inner tracking detector (ID) surrounded by a thin superconducting solenoid providing a 2\,T axial magnetic field,
electromagnetic (ECAL) and hadron (HCAL) calorimeters, and a muon spectrometer (MS).
A two-level trigger system is used to select events for storage. Events used in this analysis were selected online by single-electron or
single-muon triggers. An extensive software suite~\cite{ATL-SOFT-PUB-2021-001} is used in data simulation, in the reconstruction and analysis of real and simulated data, in detector operations, and in the trigger and data acquisition systems of the experiment.

The energy  of photon and electron candidates is reconstructed from deposits in topologically connected ECAL cells, and calibrated using information about charged-particle tracks reconstructed in the ID~\cite{EGAM-2018-01}.
Photon (electron) energy clusters are required to have a pseudorapidity in the range $|\eta|<2.37$ ($|\eta|<2.47$), excluding the transition region \mbox{$1.37<|\eta|<1.52$} between the ECAL barrel and endcaps.
Muon candidates are reconstructed~\cite{MUON-2018-03} from tracks in the MS that are matched to a corresponding track in the ID, and their pseudorapidity must satisfy $|\eta|<2.5$.
Lepton candidates must originate from the primary vertex\footnote{
The primary vertex is selected from the primary vertex candidates as the one with the highest sum of the squared transverse momenta of its associated tracks.}
and are selected by requiring $|d_{0}|/\sigma_{d_{0}} < 5.0~(3.0) $ for electrons (muons) and $|\zzsth| < 0.5$\,mm for both lepton flavors, where $d_0$ and $\sigma_{d_{0}}$ are the track's transverse impact parameter and its uncertainty, $z_0$ is the longitudinal impact parameter, and $\theta$ is the polar angle of the track's direction.
The shower shapes produced in the ECAL and HCAL along with the track information are used to identify photons and electrons.
Photon shower shapes must satisfy the \textit{Tight} photon identification criteria of \Refn{\cite{EGAM-2018-01}}.
Signal electrons must satisfy the \textit{Tight} likelihood identification criteria of \Refn{\cite{EGAM-2018-01}}, while signal muons must satisfy the \textit{Medium} identification criteria of \Refn{\cite{MUON-2018-03}}.
Photon, electron, and muon candidates are required to be isolated from other particles.
The isolation criteria limit the summed transverse momenta of tracks, and the summed transverse energies of topological clusters~\cite{PERF-2014-07}, that is allowed in separately defined conical regions around the direction of the photon or lepton.
Photon and electron candidates must satisfy the \textit{Loose} and \textit{Gradient} isolation criteria of \Refn{\cite{EGAM-2018-01}}, respectively.
Muon candidates must satisfy the \textit{PflowTight} isolation criteria of Ref.~\cite{MUON-2018-03}.
 
The neutrino's transverse momentum (\pt) is estimated from the missing transverse momentum in the event, \MET, calculated as the magnitude of the negative vector sum of the transverse momenta of all identified high-\pt\ physics objects, together with the contribution from an additional \enquote{soft term}, which is calculated from ID tracks matched to the primary vertex, but not assigned to any of the high-\pt\ objects~\cite{PERF-2016-07}.
 
The \WZy\ signal region (SR) is defined by requiring an \epem\ or \mpmm pair together with an additional $e^\pm$ or $\mu^\pm$ and at least one photon.
The three selected leptons must satisfy $\pT^\ell > 20$\,\GeV, and at least one lepton must have $\pT^\ell > 30$\,\GeV.
At least one of the electrons or muons must be matched to the trigger-level electron or muon that triggered the event.
The highest-$\pT$ photon in the event is taken as the signal photon and must have $\pTy > 15\,\GeV$.
The event is required to have $\MET > 20\,\GeV$.
To reduce the background from \ZZy\ production, the event must not contain additional leptons with $\ptl > 10$\,\GeV\ satisfying the \textit{Medium}~\cite{EGAM-2018-01,MUON-2018-03} requirement for electron and muon identification and the \textit{PflowLoose}~\cite{MUON-2018-03} requirement for muon isolation.
For the $e^+\nu_e\mu^+\mu^-\gamma$ and $\mu^+\nu_{\mu}e^+e^-\gamma$ final states, the leptons forming the $\mpmm$ or $\epem$ pair, respectively, are referred to as \enquote{\Zboson-leptons}.
For the $e^{\pm}\nu_e e^+ e^-\gamma$ and $\mu^{\pm}\nu_{\mu}\mu^+\mu^-\gamma$ final states, the leptons forming the \lplm\ pair with invariant mass closest to the nominal $\Zboson$~boson mass~\cite{Workman:2022ynf}, $\mZ$, are assigned as the \Zboson-leptons.
The third lepton remaining after assigning the \Zboson-lepton pair is called the \enquote{\Wboson-lepton}. If the \Wboson-lepton is an electron, the invariant mass of the \Wboson-lepton and the photon, \meWy, is required to satisfy $|\meWy - \mZ| > 10$\,\GeV\ to reduce the number of \ZZ\ events where one of the \Zboson\ bosons decays into an \epem\ pair and either the $e^+$ or the $e^-$ is misidentified as a photon.
The angular separation $\Delta R = \sqrt{(\Delta\eta)^2 + (\Delta\phi)^2}$ between each lepton and the photon is required to satisfy $\Delta R(\ell,\gamma) > 0.4$. This ensures reliable reconstruction of the lepton and the photon and also reduces the contribution from radiative decay (FSR) of the $\Wboson$~boson ($\Wboson\rightarrow\ell^{\pm} \nu\gamma$) and $\Zboson$~boson ($\Zboson\to\lplmy$).
The contribution from events where the photon is produced from FSR of the $\Zboson$~boson, is further reduced by requiring the invariant mass, \mll, of the \Zboson-lepton pair to exceed $81\,\GeV$.

The expected signal contribution to the selected event sample is obtained using a sample of inclusive ${\ell'}^\pm\nu\ell^{+}\ell^{-}\gamma$ signal events with invariant mass of the same-flavor opposite-charge (SFOC) lepton-pair greater than 20\,\GeV, and with the lepton--neutrino pair's invariant mass exceeding 2\,\GeV, generated by \SHERPA[2.2.11]~\cite{Bothmann:2019yzt} with the \NNPDF[3.0nnlo]~\cite{Ball:2014uwa} parton distribution function (PDF) set.
Matrix elements including all diagrams with three electroweak couplings were calculated with zero parton emissions at next-to-leading order (NLO), or with one or two partons at leading order (LO) in QCD, and at LO in the EW coupling constant
and merged with the \SHERPA parton shower~\cite{Schumann:2007mg} (PS) according to the CKKW procedure~\cite{Hoeche:2009rj}.
Photons radiated from the initial- and final-state charged particles were also generated, with a minimum photon energy requirement of 7\,\GeV\ at parton level in the matrix element calculation.

The dominant backgrounds originate from processes with a nonprompt lepton or photon from a hadron decay, or a jet misidentified as a prompt lepton or photon, e.g.~$\Zboson(\to\lplm)\gamma+X$, $t\bar{t}\gamma$, $\WZ+X$ and $\ZZ(\to\lplmlplm)+X$.
Such backgrounds are referred to as nonprompt background and estimated using data-driven techniques based on selecting event samples containing a \textit{loose lepton} and/or a \textit{loose photon} among the selected signal leptons and photon, so as to be enriched in lepton-like and/or photon-like jets.
Loose electrons must satisfy the \textit{Medium} likelihood identification requirement of \Refn{\cite{EGAM-2018-01}} and fail the \textit{Tight} identification or \textit{Gradient} isolation requirements.
Loose muons must be nonisolated and/or be matched more loosely ($3.0 < |d_0|/\sigma_{d_0} < 10.0$) than signal muons to the primary vertex.
Loose photons must fail to meet either the \textit{Loose} isolation or the \textit{Tight} identification criteria but satisfy looser ones.
The number of nonprompt background events, $N^{\textrm{nonprompt}}$, in the SR is estimated as
\begin{equation*}
N^{\textrm{nonprompt}} = \sum_{i} F_i^{\ell}(N^{\textrm{data}}_{\textrm{B},i} - N^{\textrm{prompt}}_{\textrm{B},i}) +  \sum_j F_j^{\gamma}(N^{\textrm{data}}_{\textrm{C},j} - N^{\textrm{prompt}}_{\textrm{C},j}) - \sum_{i,j}F_i^{\ell}F_j^{\gamma}(N^{\textrm{data}}_{\textrm{D},i,j} - N^{\textrm{prompt}}_{\textrm{D},i,j}),
\end{equation*}
where $F_i^{\ell}$ ($F_j^{\gamma}$) is a \textit{fake factor} defined as the ratio of the probability that a lepton-like (photon-like) jet meets the signal selection criteria to the probability that it meets the loose selection criteria, determined in bin $i$ ($j$) of the loose lepton (photon) \pT; subscripts B, C and D represent regions where events are selected with the same set of criteria as the SR but with one loose lepton, one loose photon, or one loose lepton and one loose photon, respectively; $N^{\textrm{data}}_{\textrm{X},i(j)}$ and $N^{\textrm{prompt}}_{\textrm{X},i(j)}$ represent the yields of data and of processes with prompt leptons and photons, i.e.~\WZy, \ZZy, \ZZetoy, \Zyy, in region X (X = B, C, D) and bin $i$ ($j$).
The fake factor $F^{\ell}$ is determined from a dijet event sample selected by requiring exactly one signal-lepton or one loose-lepton candidate balanced by a jet, as detailed in Ref.~\cite{STDM-2018-34}.
The fake factor $F^{\gamma}$ is determined from a $Z$+jets event sample selected by requiring a SFOC lepton-pair and either one signal-photon or one loose-photon candidate.
The yields of photon-like jet events in the signal- and loose-photon regions of the $Z$+jets sample are estimated using the data-driven method described in Ref.~\cite{STDM-2018-04}.

The \ZZy\ background contribution is estimated using a \SHERPA[2.2.11]~\cite{Bothmann:2019yzt} Monte Carlo (MC) event sample generated with the same configuration as used for the \WZy\ signal sample.
The normalisation of the \ZZy\ background is constrained using a \ZZy\ control region (CR) defined similar to the \WZy\ SR, except that the requirement on \mll\ is loosened to $\mll > 40\,\GeV$, the requirement on $\MET$ is removed,
and the veto on additional leptons is replaced by a requirement that a fourth lepton must be present with $\pT > 10\,\GeV$ satisfying looser identification and isolation criteria than for signal leptons.
 
The background from \ZZ\ events in which a \Zboson\ boson decays into an \epem\ pair and either the electron or positron is misidentified as a photon, denoted by \ZZetoy, was modeled with an inclusive sample of $pp\rightarrow\ZZ\rightarrow$\llll\ MC events generated by \POWHEGBOX~\cite{Nason:2004rx, Frixione:2007vw, Alioli:2010xd}, which was interfaced to \PYTHIA[8.210]~\cite{Sjostrand:2007gs} for parton showering and simulation of the underlying event.
The \CT[10nlo] PDF set was used for the hard-scatter process, while the \CTEQ[6L1]~\cite{Pumplin:2002vw} PDF set was used for the PS. Each reconstructed photon selected from this sample is required to be an electron in the generator's event record.
The normalisation of the \ZZetoy\ background is constrained using a CR defined similar to the \WZy\ SR, but with the \MET\ and $|\meWy-\mZ|$ criteria inverted to require $\MET<20\,\GeV$ and $|\meWy-\mZ| < 10\,\GeV$.
 
The background from \Zyy\ production where one of the photons is misidentified as an electron is estimated using an MC event sample generated with \SHERPA[2.2.10]~\cite{Bothmann:2019yzt} at NLO, with up to two additional partons at LO accuracy, and the \NNPDF[3.0nnlo]~\cite{Ball:2014uwa} set of PDFs.
 
\Pileup background, denoted by \enquote{\pileup $\gamma$}, where the photon and the trilepton system in a selected event are produced in separate \pp\ interactions, arises because the reconstructed photon's point of origin is determined relatively poorly.
This background contribution is estimated using a method similar to that introduced in \Refn{\cite{STDM-2018-04}}, where a sample of simulated \pileup events is obtained at particle level by overlaying the photon from a \yplusjets\ MC event onto an event from an inclusive $pp\rightarrow\WZ\rightarrow$\lnll\ MC sample.

The \WZy\ production cross-section is measured in a fiducial phase-space region (FR) defined at particle level by kinematic requirements closely matching those used to define the detector-level SR, using photons, electrons, muons and neutrinos in the MC event record that do not originate from the decay of a $\tau$-lepton or a hadron.
The \pt values of the three signal leptons and the neutrino must exceed 20~\GeV, and the \pt of the leading lepton must be greater than 30~\GeV.
The \pt of the signal photon must be above 15~\GeV.
A pseudorapidity requirement $|\eta^\ell| < 2.5$ ($|\eta^\gamma| < 2.37$) is imposed on leptons (photons).
The four-momenta of photons within a cone of size $\Delta R = 0.1$ around each electron or muon are added to the electron or muon four-momentum, a procedure commonly referred to as "dressing".
Each remaining prompt photon must satisfy an isolation criterion at particle level which requires the scalar sum of the \pT of all stable particles
within a cone of size $\Delta R = 0.2$ around the photon to be less than 7\% of the photon \pTy.
The angular separation between each lepton and the photon is required to satisfy $\Delta R(\ell,\gamma) > 0.4$.
The $Z$ candidate mass, \mll, must exceed 81~\GeV.
 
The \WZy\ signal event contribution in the SR is determined using a profile-likelihood fit~\cite{Cranmer:1456844} for a signal-strength parameter $\mu_{\WZy}$ which measures the signal contribution relative to the SM expectation.
The value of $\mu_{\WZy}$ is extracted simultaneously with the normalisations $\mu_{\ZZy}$ and $\mu_{\ZZ}$ of the \ZZy\ and \ZZetoy\ backgrounds, respectively, by including the dedicated \ZZy\ and \ZZetoy\ control regions in the fit. The fit is carried out for all leptonic final states combined and hence uses three bins in total: one SR and two CRs.
 
Systematic uncertainties affecting the predicted SM yields contain contributions from electron and muon triggers, reconstruction, identification~\cite{MUON-2018-03, PERF-2017-01} and isolation requirements, energy and momentum scales~\cite{MUON-2018-03, PERF-2017-03}, modeling of \MET~\cite{Aaboud:2018tkc}, and theoretical modeling of \WZy\ events. The last of these is estimated by varying the renormalisation and factorization scales, and the PDFs and \alphas, according to prescriptions in Ref.~\cite{ATL-PHYS-PUB-2016-002, Butterworth:2015oua}.
Other contributions include uncertainties from the determination of lepton and photon fake factors and modeling of prompt backgrounds in looser lepton and/or photon regions, uncertainties in the \Zyy\ cross-section and \pileup background, and signal and background uncertainties due to limited sample size. The dominant systematic uncertainty in the measured cross-section in the FR, $\sigma_{\WZy}$, arises from the data sample size in the loose lepton and/or photon region and is 5.4\%, followed by a 2.5\% uncertainty from the photon identification and isolation efficiency, and a 2.4\% uncertainty related to calibrations of muon isolation, identification, and reconstruction efficiencies, and momentum resolution and scale. The systematic uncertainties are included in the fit as nuisance parameters constrained by Gaussian probability density functions, except for statistical uncertainties of the signal and backgrounds, which are constrained by Poisson probability density functions.
 
\begin{table}[ht]
\centering
\caption{The data event yield and post-fit signal and background yields in the SR, and CRs for $ZZ\gamma$ and $ZZ(e\rightarrow\gamma)$. The uncertainties include both the statistical and systematic contributions. The uncertainty in the total yield can be smaller than the quadrature sum of the contributions because of correlations resulting from the fit.
}
\begin{tabular}{lS[table-format=3.2]@{\(\,\pm\,\)}S[table-format=2.2]S[table-format=2.3]@{\(\,\pm\,\)}S[table-format=1.3]S[table-format=2.2]@{\(\,\pm\,\)}S[table-format=1.2, table-number-alignment=left]}
\toprule
Process          & \multicolumn{2}{c}{SR}                                                   & \multicolumn{2}{c}{$ZZ\gamma$ CR}         & \multicolumn{2}{c}{$ZZ(e\rightarrow \gamma)$ CR}\\
\midrule
$WZ\gamma$      & 92 & 15              & 0.21 & 0.07  & 0.56 & 0.14  \\
$ZZ\gamma$       & 10.7 &2.3                & 23 &5 & 1.8 & 0.4  \\
$ZZ(e\rightarrow\gamma)$  &3.0 & 0.6        & 0.028 & 0.020  & 30 & 6  \\
$Z\gamma\gamma$  & 1.05 & 0.32               & 0.15 & 0.06  & 0.29 & 0.10  \\
Nonprompt background  & 30  & 6               & \multicolumn{2}{c}{-}             & \multicolumn{2}{c}{-}           \\ 
\Pileup $\gamma$ & 1.9 & 0.7                & \multicolumn{2}{c}{-}             & \multicolumn{2}{c}{-}           \\
\midrule
Total yield      & 139   & 12        & 23  & 5  & 33  & 6 \\
\midrule
Data             & \multicolumn{2}{c}{139}              & \multicolumn{2}{c}{23}        & \multicolumn{2}{c}{33} \\
\bottomrule
\end{tabular}
\label{tab:yields_postFit_Unblind}
\end{table}

\begin{figure}[h!]  
\begin{center}
\includegraphics[width=0.35\textwidth]{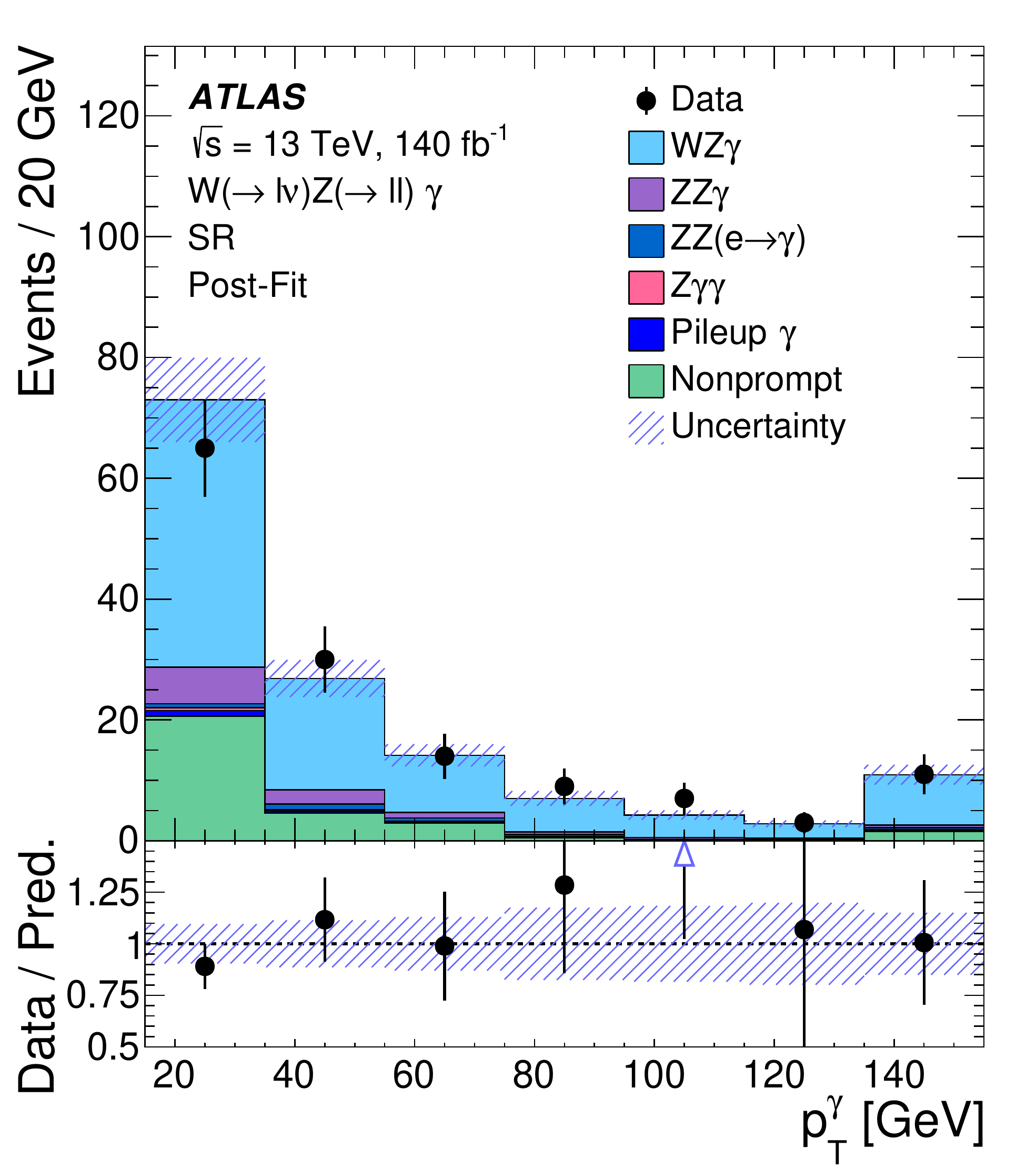}
\includegraphics[width=0.35\textwidth]{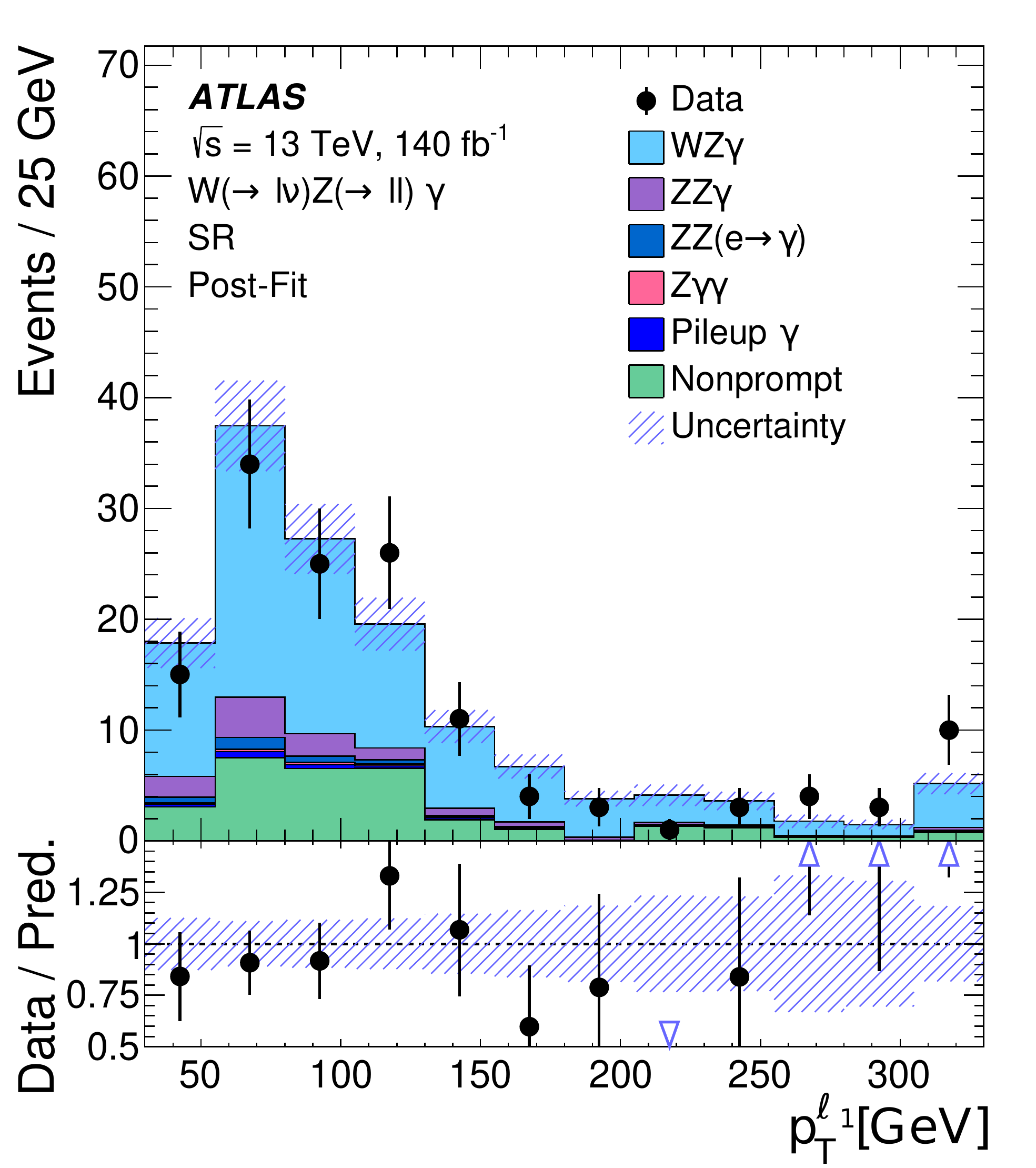} \\
\includegraphics[width=0.35\textwidth]{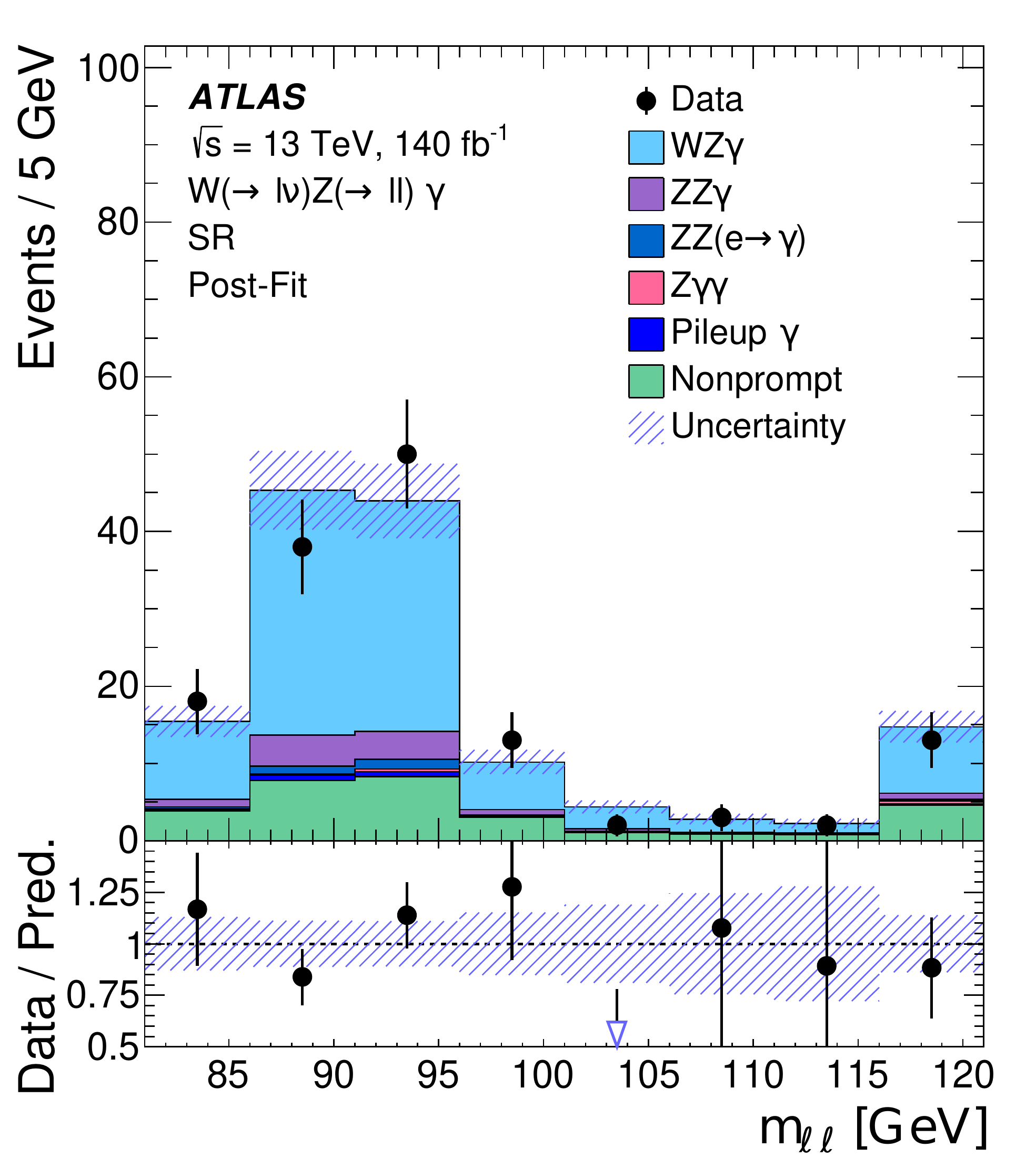}
\includegraphics[width=0.35\textwidth]{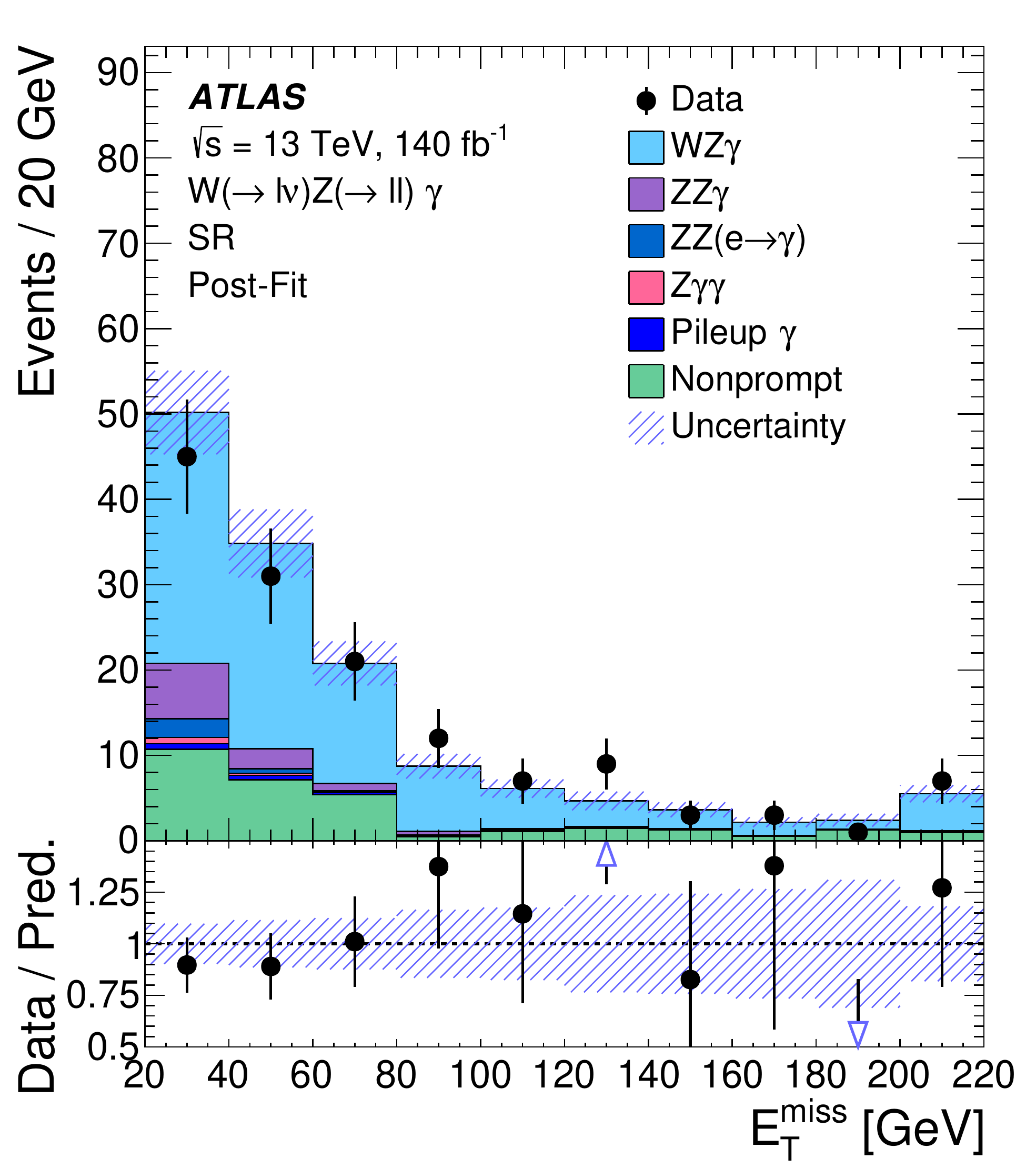}
\end{center}
\caption{Distributions of photon $p_{\mathrm{T}}^{\gamma}$ (top left), leading-lepton $p_{\mathrm{T}}^{\ell_{1}}$ (top right), \mll\ (bottom left) and \MET\ (bottom right) in the SR.
The lower panel in each figure shows the ratio of the data points to the post-fit total prediction.
The arrows indicate that the ratio lies outside the range covered by the vertical axis.
The uncertainty bands include both the statistical and systematic uncertainties as obtained by the fit.
The overflow content of each histogram is added to the last bin.
}
\label{fig:results}
\end{figure}

The background-only hypothesis is rejected with an observed (expected) significance of 6.3 (5.0) standard deviations. The observed signal strength is $\mu_{\WZy} = 1.34 \pm 0.20~\textrm{(stat.)} \pm 0.10~\textrm{(syst.)} \pm 0.07~\textrm{(theory)}$, the uncertainty being dominated by a statistical uncertainty of 15\%. The obtained $\mu_{\WZy}$ is consistent with those obtained separately from the four final states.
The fitted values of $\mu_{\ZZy}$ and $\mu_{\ZZ}$ are $1.19 \pm 0.25$ and $0.98 \pm 0.19$, respectively. The post-fit yields of the signal, backgrounds and data are shown in Table~\ref{tab:yields_postFit_Unblind}. \WZy\ events where signal electrons or muons are products of $\tau$-lepton decays, constitute 5\% of the total \WZy\ yield in the SR, and are scaled by $\mu_{\WZy}$. Figure~\ref{fig:results} compares the data with the post-fit signal and background predictions for the photon \pTy, leading-lepton \ptl, \mll, and \MET\ distributions in the SR. Good agreement is observed for all distributions.

The predicted SM fiducial cross-section, $\sigma\td{fid.}\tu{SM}$, obtained using the \SHERPA[2.2.11] event generator is $1.50 \pm 0.01~({\text{stat.}}) \pm 0.02~({\text{PDF+}}\alphas) \pm 0.06~({\text{scale}})~\fb$.
This value does not include the effect of NLO EW corrections, which has been found to be $K_\mathrm{EW} = \sigma\td{fid.}\tu{NLO EW}/\sigma\td{fid.}\tu{LO} = 1.05$~\cite{cheng2021nlo} for the subprocess $\pp\to WZ\gamma \to e^+\nu_e\mu^+\mu^-\gamma$.
The measured cross-section in the FR is $\sigma_{\WZy} = \mu_{\WZy} \cdot \sigma\td{fid.}\tu{SM} = 2.01 \pm 0.30~({\text{stat.}}) \pm 0.16~({\text{syst.}})~\fb$, which is consistent with the SM prediction to within 1.5~standard deviations.

In conclusion, the process $pp\rightarrow WZ\gamma$ has been observed by the ATLAS
detector at the LHC. Events with three prompt leptons, containing one same-flavor opposite-charge pair, plus one prompt photon and missing transverse momentum were selected from a 140~fb$^{-1}$ data set collected from $\sqrt{s} = 13$~\TeV\ proton--proton collisions. The background-only hypothesis is rejected with an observed (expected) significance of 6.3 (5.0) standard deviations. The $\pp\to WZ\gamma \to{\ell'}^{\pm}\nu\ell^+\ell^-\gamma$ ($\lorlprime = e, \mu$) cross-section in the fiducial phase space defined by kinematic requirements on the \lnlly\ system and by isolation requirements on the photon is measured to be $2.01 \pm 0.34~\textrm{fb}$.

\section*{Acknowledgments}

% The next lines are included from the .//acknowledgements/Acknowledgements.tex input file

We thank CERN for the very successful operation of the LHC, as well as the
support staff from our institutions without whom ATLAS could not be
operated efficiently.
 
We acknowledge the support of
ANPCyT, Argentina;
YerPhI, Armenia;
ARC, Australia;
BMWFW and FWF, Austria;
ANAS, Azerbaijan;
CNPq and FAPESP, Brazil;
NSERC, NRC and CFI, Canada;
CERN;
ANID, Chile;
CAS, MOST and NSFC, China;
Minciencias, Colombia;
MEYS CR, Czech Republic;
DNRF and DNSRC, Denmark;
IN2P3-CNRS and CEA-DRF/IRFU, France;
SRNSFG, Georgia;
BMBF, HGF and MPG, Germany;
GSRI, Greece;
RGC and Hong Kong SAR, China;
ISF and Benoziyo Center, Israel;
INFN, Italy;
MEXT and JSPS, Japan;
CNRST, Morocco;
NWO, Netherlands;
RCN, Norway;
MEiN, Poland;
FCT, Portugal;
MNE/IFA, Romania;
MESTD, Serbia;
MSSR, Slovakia;
ARRS and MIZ\v{S}, Slovenia;
DSI/NRF, South Africa;
MICINN, Spain;
SRC and Wallenberg Foundation, Sweden;
SERI, SNSF and Cantons of Bern and Geneva, Switzerland;
MOST, Taiwan;
TENMAK, T\"urkiye;
STFC, United Kingdom;
DOE and NSF, United States of America.
In addition, individual groups and members have received support from
BCKDF, CANARIE, Compute Canada and CRC, Canada;
PRIMUS 21/SCI/017 and UNCE SCI/013, Czech Republic;
COST, ERC, ERDF, Horizon 2020 and Marie Sk{\l}odowska-Curie Actions, European Union;
Investissements d'Avenir Labex, Investissements d'Avenir Idex and ANR, France;
DFG and AvH Foundation, Germany;
Herakleitos, Thales and Aristeia programmes co-financed by EU-ESF and the Greek NSRF, Greece;
BSF-NSF and MINERVA, Israel;
Norwegian Financial Mechanism 2014-2021, Norway;
NCN and NAWA, Poland;
La Caixa Banking Foundation, CERCA Programme Generalitat de Catalunya and PROMETEO and GenT Programmes Generalitat Valenciana, Spain;
G\"{o}ran Gustafssons Stiftelse, Sweden;
The Royal Society and Leverhulme Trust, United Kingdom.
 
The crucial computing support from all WLCG partners is acknowledged gratefully, in particular from CERN, the ATLAS Tier-1 facilities at TRIUMF (Canada), NDGF (Denmark, Norway, Sweden), CC-IN2P3 (France), KIT/GridKA (Germany), INFN-CNAF (Italy), NL-T1 (Netherlands), PIC (Spain), ASGC (Taiwan), RAL (UK) and BNL (USA), the Tier-2 facilities worldwide and large non-WLCG resource providers. Major contributors of computing resources are listed in Ref.~\cite{ATL-SOFT-PUB-2023-001}.

% End of text imported from the .//acknowledgements/Acknowledgements.tex input file

\printbibliography
 
\clearpage
% ATLAS Collaboration author list
% Reference date of STDM-2019-17 is 2023-03-03
% Author list last updated on date 17-DEC-23
% Data extracted on 17-Dec-2023 for paper reference STDM-2019-17
% at 12:19pm
 
\begin{flushleft}
\hypersetup{urlcolor=black}
{\Large The ATLAS Collaboration}

\bigskip

\AtlasOrcid[0000-0002-6665-4934]{G.~Aad}$^\textrm{\scriptsize 102}$,
\AtlasOrcid[0000-0002-5888-2734]{B.~Abbott}$^\textrm{\scriptsize 120}$,
\AtlasOrcid[0000-0002-1002-1652]{K.~Abeling}$^\textrm{\scriptsize 55}$,
\AtlasOrcid[0000-0001-5763-2760]{N.J.~Abicht}$^\textrm{\scriptsize 49}$,
\AtlasOrcid[0000-0002-8496-9294]{S.H.~Abidi}$^\textrm{\scriptsize 29}$,
\AtlasOrcid[0000-0002-9987-2292]{A.~Aboulhorma}$^\textrm{\scriptsize 35e}$,
\AtlasOrcid[0000-0001-5329-6640]{H.~Abramowicz}$^\textrm{\scriptsize 151}$,
\AtlasOrcid[0000-0002-1599-2896]{H.~Abreu}$^\textrm{\scriptsize 150}$,
\AtlasOrcid[0000-0003-0403-3697]{Y.~Abulaiti}$^\textrm{\scriptsize 117}$,
\AtlasOrcid[0000-0003-0762-7204]{A.C.~Abusleme~Hoffman}$^\textrm{\scriptsize 137a}$
% ...
\bigskip
\\

% ...
$^{29}$Physics Department, Brookhaven National Laboratory, Upton NY; United States of America.\\
% ...
$^{35}$$^{(a)}$Facult\'e des Sciences Ain Chock, R\'eseau Universitaire de Physique des Hautes Energies - Universit\'e Hassan II, Casablanca;$^{(b)}$Facult\'{e} des Sciences, Universit\'{e} Ibn-Tofail, K\'{e}nitra;$^{(c)}$Facult\'e des Sciences Semlalia, Universit\'e Cadi Ayyad, LPHEA-Marrakech;$^{(d)}$LPMR, Facult\'e des Sciences, Universit\'e Mohamed Premier, Oujda;$^{(e)}$Facult\'e des sciences, Universit\'e Mohammed V, Rabat;$^{(f)}$Institute of Applied Physics, Mohammed VI Polytechnic University, Ben Guerir; Morocco.\\
% ...
$^{49}$Fakult\"{a}t Physik , Technische Universit{\"a}t Dortmund, Dortmund; Germany.\\
% ...
$^{55}$II. Physikalisches Institut, Georg-August-Universit\"{a}t G\"ottingen, G\"ottingen; Germany.\\
% ...
$^{102}$CPPM, Aix-Marseille Universit\'e, CNRS/IN2P3, Marseille; France.\\
% ...
$^{117}$Department of Physics, New York University, New York NY; United States of America.\\
% ...
$^{120}$Homer L. Dodge Department of Physics and Astronomy, University of Oklahoma, Norman OK; United States of America.\\
% ...
$^{137}$$^{(a)}$Departamento de F\'isica, Pontificia Universidad Cat\'olica de Chile, Santiago;$^{(b)}$Millennium Institute for Subatomic physics at high energy frontier (SAPHIR), Santiago;$^{(c)}$Instituto de Investigaci\'on Multidisciplinario en Ciencia y Tecnolog\'ia, y Departamento de F\'isica, Universidad de La Serena;$^{(d)}$Universidad Andres Bello, Department of Physics, Santiago;$^{(e)}$Instituto de Alta Investigaci\'on, Universidad de Tarapac\'a, Arica;$^{(f)}$Departamento de F\'isica, Universidad T\'ecnica Federico Santa Mar\'ia, Valpara\'iso; Chile.\\
% ...
$^{150}$Department of Physics, Technion, Israel Institute of Technology, Haifa; Israel.\\
$^{151}$Raymond and Beverly Sackler School of Physics and Astronomy, Tel Aviv University, Tel Aviv; Israel.\\
% ...
$^{a}$ Also Affiliated with an institute covered by a cooperation agreement with CERN.\\
$^{b}$ Also at An-Najah National University, Nablus; Palestine.\\
$^{c}$ Also at Borough of Manhattan Community College, City University of New York, New York NY; United States of America.\\
$^{d}$ Also at Center for High Energy Physics, Peking University; China.\\
$^{e}$ Also at Center for Interdisciplinary Research and Innovation (CIRI-AUTH), Thessaloniki; Greece.\\
$^{f}$ Also at Centro Studi e Ricerche Enrico Fermi; Italy.\\
% ...

\end{flushleft}

% Created with Glance <Atlas.Glance@cern.ch>

\clearpage
 
\end{document}